\documentclass[aps,prd,secnumarabic,nobalancelastpage,amsmath,amssymb,
nofootinbib,twocolumn,showpacs]{revtex4-1}

\newcommand\be{\begin{equation}}
\newcommand\ee{\end{equation}}
\newcommand\nono{\nonumber}
\newcommand\bse{\begin{subequations}}
\newcommand\ese{\end{subequations}}
\newcommand\bea{\begin{eqnarray}}
\newcommand\eea{\end{eqnarray}}
\newcommand{\Zdown}[2]{Z^{#1}\! ,_{#2}}
\newcommand{\Zup}[2]{Z^{#1}\! ,^{#2}}
\newcommand{\Xup}[2]{X^{#1}\!,_{#2}}

\usepackage[toc,page]{appendix}
\usepackage{graphics}      
\usepackage{graphicx}      
\usepackage{longtable}     
\usepackage{url}           
\usepackage{bm}            
\usepackage{mathrsfs}
\usepackage{hyperref}
\usepackage{color}
\usepackage{soul}
\usepackage{tikz}
\usetikzlibrary{calc}
\usetikzlibrary{arrows.meta,arrows}
\usepackage{tikz-3dplot}

\numberwithin{equation}{section}

\newcommand\ringring[1]{%
  {
   \mathop{\kern0pt #1}\limits^{
     \vbox to-1.85ex{
       \kern-2ex 
       \hbox to 0pt{\hss\normalfont\kern.1em \r{}\kern-.45em \r{}\hss}%
       \vss 
     }
   }
  }
}
\begin{document}
\title{Elasticity Theory in General Relativity}
\author{J. David Brown}
\email{david{\_}brown@ncsu.edu}
\affiliation{Department of Physics, North Carolina State University, Raleigh, NC 27695}
\date{\today}

\begin{abstract}
The general relativistic theory of elasticity is reviewed from a Lagrangian, as opposed to Eulerian, perspective. The equations 
of motion and stress--energy--momentum tensor for a hyperelastic body are derived from the gauge--invariant action principle first considered 
by DeWitt. This action is  a natural extension of the action  for a single relativistic particle. The central object in the Lagrangian treatment is the 
Landau--Lifshitz radar metric, which is the relativistic version of the right Cauchy--Green deformation tensor. We also introduce relativistic 
definitions of the deformation gradient, Green strain, and first and second Piola--Kirchhoff stress tensors. 
A gauge--fixed description of relativistic hyperelasticity is also presented, and the nonrelativistic theory is derived in the limit as the 
speed of light becomes infinite.
\end{abstract}
\maketitle
\makeatletter
\let\toc@pre\relax
\let\toc@post\relax
\makeatother


\section{Introduction}\label{introduction} 
Elasticity theory in the nonrelativistic regime is a well--developed subject,
with applications to many branches of engineering and science. (See, for example, 
Refs.~\cite{Bower,BonetWood,Kelly,Hackett}.)
The special relativistic theory dates back to Herglotz \cite{Herglotz}. 
This work was extended to general relativity by DeWitt \cite{DeWitt62}, who tied extra degrees of freedom (a framework of ``clocks") to the 
material elements. The elastic material with clocks provides
a natural coordinate system that DeWitt used to investigate a quantum theory of gravity. 
Some recent works on general relativistic elasticity include Carter and Qunitana \cite{CarterQuintana}, Kijowski and Magli \cite{KijowskiMagli}, 
Marsden and Hughes \cite{MarsdenHughes}, 
Beig and Schmidt \cite{Beig:2002pk}, Beig and Wernig--Pichler \cite{Beig:2005hv}, 
Gundlach, Hawke and Erickson \cite{Gundlach:2011nt}, Andersson \cite{Andersson:2014sga} and Andersson, Oliynyk and Schmidt \cite{Andersson:2014aza}.  
See also the extensive treatment by Wernig--Pichler \cite{Wernig-Pichler}.

Elastic materials differ from fluids in that they allow for the presence of shear stresses. In most astrophysical contexts, shear stresses are negligible.
The material in most stars, jets, accretion disks and large planets are all well described as fluids. The interstellar and intergalactic media, and 
matter on cosmological scales, are modeled as fluids.  One 
astrophysical context in which shear stress is important is the crust of a neutron star \cite{ChamelHaensel}. 
This is the main practical motivation for the development of 
a relativistic theory of elasticity. Of course, intellectual curiosity also serves as motivation. Can we compute, from first principles, the behavior 
of a rubber ball as it falls into a black hole? 

An elastic body is described mathematically in terms of ``matter space", whose points coincide with the material elements (or particles) that make up the body. 
Coordinates on matter space serve as labels for the elements. The motion of the body through spacetime can be described using either 
the ``Eulerian" or ``Lagrangian" perspective. In the Eulerian approach, the independent variables of the theory are events in spacetime. 
The dependent variables are the matter space coordinates for the material element whose worldline passes through that event. 
With the Lagrangian approach, the independent variables are the matter space coordinates for a given  element of the body, and a worldline parameter. 
The dependent variables are spacetime events.

The original works of Herglotz and DeWitt used the Lagrangian perspective. 
Many of the more recent studies of relativistic elasticity have  focused 
 on the Eulerian perspective~\cite{CarterQuintana,KijowskiMagli,Beig:2002pk,Gundlach:2011nt,Andersson:2014sga}.  Notable exceptions
are found in Refs.~\cite{Beig:2005hv,Andersson:2014aza}, where the Lagrangian description is used to address the existence and uniqueness of 
solutions to elastic body motion in general relativity. 

This paper presents a detailed account of relativistic elasticity theory from the Lagrangian perspective. 
The approach is not mathematically rigorous  and not mathematically sophisticated;  basic tensor notation is used throughout. 
Many of the results can be found scattered throughout the literature 
\cite{CarterQuintana,KijowskiMagli,MarsdenHughes,Beig:2002pk,Beig:2005hv,Gundlach:2011nt,Andersson:2014sga}, although
these results can be difficult 
to recognize due to the variety of notations and the varying levels of mathematical rigor used by different authors.  The goal here is 
to provide a comprehensive overview of the subject that is accessible to a wide audience. 

One advantage of the 
Lagrangian approach is that it easily incorporates ``natural" boundary conditions in which the surface of the body is free from physical constraints  
or forces. The  Lagrangian description can also be more efficient computationally for a finite--size body, since the independent variables are the 
material points rather than the entire spacetime.  A possible disadvantage of the Lagrangian description, as compared to the Eulerian description,  
is that it might be more difficult to treat material shocks and discontinuities. 

In Sec.~(\ref{kinematics}), we establish notation and introduce the kinematical relationships needed to describe relativistic elastic materials. 
Central to the description of elastic materials is the Landau--Lifshitz radar metric, discussed in Sec.~(\ref{radarsection}). The radar metric defines the proper distance between neighboring material elements as measured in the local rest frame of an element. 
In Sec.~(\ref{strain}) we identify the radar metric as the right Cauchy--Green deformation tensor and define the Lagrangian (or Green) strain 
tensor. The action principle for a relativistic hyperelastic body is presented in Sec.~(\ref{actionandeoms}). The action principle determines 
the bulk equations  of motion as well as  the natural boundary conditions for the body. Section (\ref{SEMsection}) contains a detailed calculation of the 
stress--energy--momentum (SEM) tensor for the hyperelastic body. The spatial components of the SEM tensor are given by the second Piola--Kirchhoff 
stress tensor for the material. In Sec.~(\ref{ppsection}) we review the relativistic point particle. We begin the analysis using an arbitrary worldline parameter, 
then transform to the ``gauge fixed" description by tying the parameter to a spacetime coordinate. Section~(\ref{gaugesection}) repeats the analysis 
for the hyperelastic model, arriving at  gauge fixed forms for the action principle and equations of motion. In Sec.~(\ref{NRsection}) we obtain the nonrelativistic limit 
of the elastic body action and equations of motion by taking the speed of light to infinity. Section~(\ref{modelsection}) contains a discussion of various 
constitutive models for hyperelastic materials. Models are specified by giving the energy density as a function of either the Lagrangian strain, the 
stress invariants, or the principal stretches of the material. 

The sign conventions of Misner, Thorne and Wheeler \cite{MTW} are used throughout. 

\section{Kinematics} \label{kinematics}
Let $x^\mu$  denote the spacetime coordinates and  $g_{\mu\nu}$ denote the spacetime metric. Indices on spacetime tensors are lowered and raised with  
$g_{\mu\nu}$ and it's inverse $g^{\mu\nu}$, as usual. 

A continuous body is a congruence of worldlines defined by $x^\mu = X^\mu(\lambda,\zeta)$ 
where the parameters $\zeta^i$ (with $i = 1,2,3$) label the continuum of material ``particles" (or points, or elements) in the body and $\lambda$ parametrizes the 
worldline of each particle. Typically the functions $X^\mu(\lambda,\zeta)$ are only defined for finite ranges of the labels $\zeta^i$. 
Correspondingly, the worldlines do not always fill the entire spacetime, but rather a subset of spacetime, the body's ``world tube". 
The space of material particles (or elements) is called ``matter space," denoted ${\cal S}$. The labels $\zeta^i$ serve as coordinates in ${\cal S}$. 
\begin{figure}[b]
\centering
\begin{tikzpicture}
\begin{scope} [ ]
\draw plot [smooth cycle, tension=0.8] coordinates {(-1.3,-0.5)  (1.4,-0.5)  (1.1,1.0)  (-1.1,1.0)};
\draw [->] (0,0)--(-0.5,-0.4) node [left] {$\zeta^1$}; 
\draw [->] (0,0)--(1.0,0) node [right] {$\zeta^2$}; 
\draw [->] (0,0)--(0,1.0) node [left] {$\zeta^3$}; 
\node at (-1.2,0.4) {${\cal S}$};
\node (dot) at (0.5,0.5) {};
\draw [fill] (dot) circle (0.04);
\end{scope}
\begin{scope}[ shift={(3.5,0)} ]
\draw plot [smooth cycle, tension=0.3] coordinates { (-1.1,-0.8) (1.4, -0.8) (1.4,2.3) (-1.1,2.3) };
\draw [->] (0,0)--(-0.5,-0.4) node [left] {$x^1$}; 
\draw [->] (0,0)--(1.0,0) node [right] {$x^2$}; 
\draw [->] (0,0)--(0,1.0) node [left] {$x^0$}; 
\draw [->] plot [smooth] coordinates { (0.5,-0.5) (0.65,0.25) (0.35,1.25)  (0.5,2.2) };
\node at (-0.7,2.1) {${\cal M}$};
\end{scope}
\draw [-{Stealth[scale=1.3,angle'=45]}] plot [smooth, tension=0.6] coordinates {(dot)  (1.3, 1.3) (2.3,1.6) (3.88, 1.5)}; 
\node at (1.2,1.8) {${X^\mu(\lambda,\zeta)}$}; 
\end{tikzpicture}
\caption{ $X^\mu(\lambda,\zeta)$ with $\lambda \in \Re$ maps the point with coordinates $\zeta^i$ in the three--dimensional matter space 
${\cal S}$ to a worldline in the four--dimensional spacetime ${\cal M}$.} 
\label{kinematicsfig}
\end{figure}
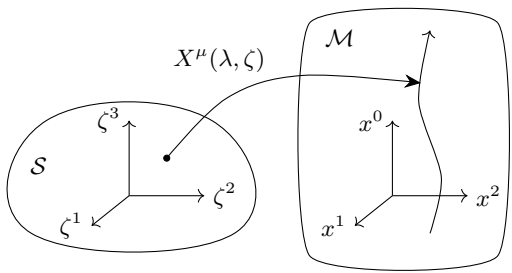

The functions $X^\mu(\lambda,\zeta)$ constitute a mapping from $\Re \times {\cal S}$ to spacetime ${\cal M}$. That is, for each point $\zeta^i$ in matter 
space, $X^\mu(\lambda,\zeta)$ sweeps out a timelike worldline in spacetime as the parameter $\lambda$ ranges over real values; 
see Fig.~(\ref{kinematicsfig}).  The 
inverse mappings are defined by $\lambda = \Lambda(x)$ and $\zeta^i = Z^i(x)$. Thus, given a spacetime event $x^\mu$ inside the world tube of the body, 
$\Lambda$ gives the parameter value $\lambda$ and $Z^i$ give the labels $\zeta^i$ of the point in the body that passes through that event. We therefore have the identities 
$\lambda = \Lambda(X(\lambda,\zeta))$ and $\zeta^i = Z^i(X(\lambda,\zeta))$, and differentiation with respect to $\lambda$ and $\zeta^i$ yields 
\bse\label{orthorelations}\bea
	\Xup{\mu}{i} \Zdown{j}{\mu} & = & \delta_i^j \ ,\\
	\dot X^\mu \Lambda,_\mu & = & 1 \ , \\
	\Xup{\mu}{i} \Lambda,_\mu & = & 0 \ ,\\
	\dot X^\mu \Zdown{i}{\mu} & = & 0 \ .
\eea\ese
Here, ``$,\mu$" denotes $\partial/\partial x^\mu$, ``$,i$" denotes $\partial/\partial \zeta^i$, and the overdot denotes $\partial/\partial\lambda$. 
We can also use the identity $x^\mu = X^\mu(\Lambda(x), Z(x))$  to derive the  useful relation 
\be\label{invorthorelation}
	 \dot X^\mu \Lambda,_\nu + \Xup{\mu}{i} \Zdown{i}{\nu} = \delta^\mu_\nu \ ,
\ee
by differentiating  with respect to the spacetime coordinates $x^\nu$ of events inside the body's world tube.

The velocities of the material particles are defined by 
\be\label{velocitydef}
	U^\mu  =  \dot X^\mu / \alpha  
\ee
where 
\be\label{alphadef}
	\alpha = \sqrt{-\dot X^\mu \dot X_\mu } \ ,
\ee
is  the {\em material lapse function}. That is, 
$\alpha \, d\lambda$  is the interval of proper time along the material worldlines between $\lambda$ and $\lambda + d\lambda$. 
Equation (\ref{orthorelations}d) tells us that the covectors  $\Zdown{i}{\mu}$ are orthogonal to $U^\mu$:
\be
	U^\mu \Zdown{i}{\mu} = 0 \ .
\ee
Thus, the vectors $\Zup{i}{\mu} \equiv g^{\mu\nu} \Zdown{i}{\nu}$  are purely spatial as viewed in the rest frame of the material. That is, along 
a given material particle worldline, the vectors  $\Zup{i}{\mu}$
span the set of nearby events that are seen as simultaneous by an observer moving along that worldline. 

\section{Radar Metric}\label{radarsection}
The {\em radar metric} is defined inside the world tube of the body by
\be\label{radarmetric}
	f_{\mu\nu} = g_{\mu\nu} + U_\mu U_\nu  \ .
\ee
This is sometimes called the {\em Lagrangian metric} \cite{DeWitt62}. 
The name radar metric stems from the analysis of  Landau and Lifshitz \cite{LandauLifshitz}, who show that this tensor defines the distance 
between nearby objects  by reflecting 
light rays between the objects' worldlines.\footnote{Landau and Lifshitz use a spacetime coordinate system in which the spatial coordinates $x^a$ are tied to 
the worldlines. Then the worldline velocity has components $U^a = 0$ and $U^0 =1/\sqrt{-g_{00}}$. It follows that the spatial components of 
the radar metric  are $f_{ab} = g_{ab} - g_{0a} g_{0b}/g_{00}$, where $a$ and $b$ are spatial indices.}

The radar metric satisfies $f_{\mu\nu} U^\nu = 0$. It acts as a projection operator that projects tensors into the subspace orthogonal to the worldlines. 
For example, given a vector $V^\mu$  inside the body's world tube,  the vector $f^\mu_\nu V^\nu$ is orthogonal to the worldlines. Likewise, 
for any covector $W_\mu$ inside the body's world tube, the covector $f_\mu^\nu W_\nu$ is orthogonal to the worldlines. 
A tensor is called ``spatial" if it is orthogonal to the worldlines in each of its indices. If $V^\mu$ is a spatial vector, it satisfies $V^\mu = f^\mu_\nu V^\nu$. If 
$W_\mu$ is a spatial covector, it satisfies $W_\mu = f^\nu_\mu W_\nu$. 

The spacetime metric $g_{\mu\nu}$ defines the inner product between vectors. It also determines the 
spacelike or timelike separation between neighboring events, as follows. Consider a vector $V^\mu$ at some event ${\cal P}$. We 
can construct a parametrized curve $X^\mu_{crv}(\sigma) = V^\mu\sigma$ 
that passes through ${\cal P}$ at $\sigma=0$. The tangent to the curve at ${\cal P}$ is $V^\mu = \partial X^\mu_{crv}(0)/\partial \sigma$. 
The coordinate separation between ${\cal P}$ (at $\sigma=0$) and the nearby event
$\sigma=d\sigma$ is the ``separation vector" $dx^\mu \equiv X^\mu_{crv}(d\sigma) - X^\mu_{crv}(0) = V^\mu d\sigma$. 
The magnitude of the separation vector, given by the inner  product of $V^\mu d\sigma$ with itself, 
defines the proper distance $ds$ between events: $ds^2 \equiv |V d\sigma|^2 = g_{\mu\nu} V^\mu V^\nu d\sigma^2 
= g_{\mu\nu} dx^\mu dx^\nu$. 

If the separation vector $dx^\mu = V^\mu d\sigma$ at ${\cal P}$ is a spatial vector, then the neighboring events are simultaneous as seen by the observer 
who is at rest with the material element that passes through ${\cal P}$. For these events, the spacelike separation is given by
$ds^2 = f_{\mu\nu} dx^\mu dx^\nu$. That is, at each event ${\cal P}$ within the body's world tube, the radar metric determines proper distances 
within the subspace  orthogonal to the worldline passing through ${\cal P}$. 

The radar metric also acts as an inner product: $f_{\mu\nu} V^\mu W^\nu$. If $V^\mu$ and $W^\nu$ are non--spatial vectors, the radar metric 
eliminates the non--spatial components. Thus, the result $f_{\mu\nu} V^\mu W^\nu = (f^\mu_\alpha W^\alpha) g_{\mu\nu} (f^\nu_\beta W^\beta)$ shows 
that $f_{\mu\nu} V^\mu W^\nu$ coincides with the inner product between spatial vectors $f^\mu_\nu V^\nu$ and $f^\mu_\nu W^\nu$, as defined by 
the spacetime metric. 

Any spacetime tensor defined at a point in the world tube of the body can be mapped into the matter space ${\cal S}$ using $Z^i_{,\mu}$ for contravariant 
indices and $X^\mu_{,i}$ for covariant indices. We denote this process by replacing Greek indices with Latin indices. For example, the 
spacetime vector $V^\mu$ is mapped to ${\cal S}$ by $V^i \equiv V^\mu Z^i_{,\mu}$. The spacetime covector $W_\mu$ is mapped to ${\cal S}$ by 
$W_i \equiv W_\mu X^\mu_{,i}$. An important example is the radar metric and its inverse: 
\bse\label{radarmetriconS}\bea
	f_{ij} & = & X^\mu_{,i} f_{\mu\nu} X^\nu_{,j} \ ,\\
	f^{ij} & = & Z^i_{,\mu} f^{\mu\nu} Z^j_{,\nu} \ .
\eea\ese
Using Eqs.~(\ref{orthorelations}) and (\ref{invorthorelation}), one can verify that $f^{ij}$ is the inverse of $f_{ij}$. It is useful to note that $f^{ij} = Z^i_{,\mu} g^{\mu\nu} Z^j_{,\nu}$; however, $f_{ij} \ne X^\mu_{,i} g_{\mu\nu} X^\nu_{,j} $.  

If $W_\mu$ is a spatial covector, then the definition $W_i \equiv W_\mu \Xup{\mu}{i}$ can be inverted  by writing $W_\mu = W_i\Zdown{i}{\mu}$. 
To verify this result, we use Eq.~(\ref{invorthorelation}) and the fact that $\dot X^\mu$ and $W_\mu$ are orthogonal. Note, however, that for a 
spatial vector $V^\mu$, the definition $V^i \equiv V^\mu \Zdown{i}{\mu}$ is {\em not} inverted in a similar way: $V^\mu \ne V^i \Xup{\mu}{i}$.  
In particular, we have 
\bse\bea
	f_{\mu\nu} & = & f_{ij} \Zdown{i}{\mu} \Zdown{j}{\nu} \ ,\\
	f^{\mu\nu} & \ne & f^{ij} \Xup{\mu}{i} \Xup{\nu}{j} 
\eea\ese
for the radar metric.

Generally, the mapping of tensors to ${\cal S}$ is reserved for spatial tensors. For spatial tensors, the mapping preserves the raising and lowering 
of indices. For example, consider $V^i  \equiv \Zdown{i}{\mu} V^\mu$ and $V_i \equiv  \Xup{\mu}{i} V_\mu = \Xup{\mu}{i}
g_{\mu\nu} V^\nu$. If $V^\mu$ is a spatial vector, we can verify that $V_i = f_{ij} V^j$ and $V^i = f^{ij} V_j$. 
On the other hand, if $V^\mu$ is not spatial, the raising and lowering 
of indices is not preserved. Consider, for example, the material velocity $U^\mu$, which of course is not spatial. We have $U^i \equiv U^\mu \Zdown{i}{\mu} = 0$
and $U_i \equiv U_\mu \Xup{\mu}{i}$ which, in general, is not zero. Clearly  $U_i \ne f_{ij} U^j$ and $U^i \ne f^{ij} U_j$.  

The combination  $U_\mu \Xup{\mu}{i}$ appears sufficiently often that a shorthand notation is useful: 
\be
	v_i \equiv U_\mu \Xup{\mu}{i} \ ,
\ee
These are the components of the spacetime velocity of the material, as a covector, projected into the subspace $\lambda = {\rm const}$ and expressed in the coordinate system
supplied by the matter space labels $\zeta^i$. 

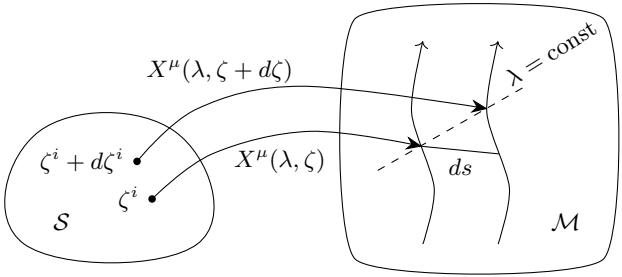
\begin{figure}[htb]
\centering
\begin{tikzpicture}
\begin{scope} [ ]
\draw plot [smooth cycle, tension=0.8] coordinates {(-1.0,-0.5)  (1.3,-0.5)  (1.0,1.0)  (-0.8,1.0)};
\node at (-0.5,-0.2) {${\cal S}$};
\node (dot) at (0.5,0.6) [circle, fill, inner sep=0pt, minimum size=0.1cm, label=left:$\zeta^i + d\zeta^i$] {};
\node (dot2) at (0.7,0.1) [circle, fill, inner sep=0pt, minimum size=0.1cm, label=left:$\zeta^i$] {};
\end{scope}
\begin{scope}[ shift={(4.5,0)} ]
\draw plot [smooth cycle, tension=0.3] coordinates { (-1.1,-0.7) (2.2, -0.7) (2.2,2.5) (-1.1,2.5) };
\draw [->] plot [smooth] coordinates { (0.8,-0.5) (0.95,0.25) (0.65,1.25)  (0.8,2.2) };
\draw [->] plot [smooth] coordinates { (-0.2,-0.5) (-0.05,0.25) (-0.35,1.25)  (-0.2,2.2) };
\node at (1.7,-0.2) {${\cal M}$};
\end{scope}
\node (endpoint) at (5.16, 1.3) {};
\node (endpoint2) at (4.29, 0.8) {};
\draw [-{Stealth[scale=1.3,angle'=45]}] plot [smooth, tension=0.6] coordinates {(dot)  (1.3, 1.3) (2.7,1.6) (endpoint)}; 
\draw [-{Stealth[scale=1.3,angle'=45]}] plot [smooth, tension=0.6] coordinates {(dot2)  (1.5, 0.7) (2.8,1.0) (endpoint2)};
\node at (1.6,1.8) {$X^\mu(\lambda,\zeta + d\zeta)$}; 
\node at (2.4,0.6) {$X^\mu(\lambda,\zeta)$}; 
\draw  (4.29,0.8) -- (5.31,0.7) node [below, midway] {$ds$};
\draw [dashed] plot [smooth] coordinates { (3.7,0.48) (5.7,1.62) };
\node [rotate=32] at (6.0,2.05) {$\lambda = {\rm const}$};
\end{tikzpicture}
\caption{ The radar metric at the event  $X^\mu(\lambda,\zeta)$ defines the proper distance $ds$ between neighboring particles 
$\zeta^i$ and $\zeta^i + d\zeta^i$ as measured in the rest 
frame of the material. The interval labeled $ds$  is orthogonal to the worldline. 
In general, the surfaces $\lambda = {\rm const}$ are not orthogonal to the worldline.} 
\label{kinematicsfig2}
\end{figure}

Consider the separation vector $dx^\mu = X^\mu(\lambda,\zeta + d\zeta) - X^\mu(\lambda,\zeta) = \Xup{\mu}{i} d\zeta^i$ connecting nearby events 
$X^\mu(\lambda,\zeta)$ and $X^\mu(\lambda,\zeta + d\zeta)$. This separation vector is not spatial. However, as with any vector, we can construct its
spatial component using the radar metric: $f^\mu_\nu dx^\mu$. The magnitude of the spatial component of the separation vector is
\bea
	ds^2 & = &  g_{\mu\nu}  (f^\mu_\alpha dx^\alpha) (f^\nu_\beta dx^\beta) = f_{\mu\nu} \Xup{\mu}{i} \Xup{\nu}{j} d\zeta^i d\zeta^j \nono\\
	& = & f_{ij} d\zeta^i d\zeta^j \ .
\eea
Thus, the radar metric on matter space,  $f_{ij}$,  defines the proper distance between the nearby events defined by projecting 
$X^\mu(\lambda,\zeta)$ and $X^\mu(\lambda,\zeta + d\zeta)$ into the subspace orthogonal to the particle worldline. That is, 
$ds^2 =  f_{ij} d\zeta^i d\zeta^j $ is the (square of the) proper distance between the worldlines of the material particles $\zeta^i$ and $\zeta^i + d\zeta^i$, as 
seen in the rest frame of the material. See Fig.~(\ref{kinematicsfig2}). 

\section{Strain}\label{strain}
In continuum mechanics \cite{Bower,BonetWood,Kelly,Hackett},  the strain of the material is quantified by the  {\em deformation gradient}. 
In the relativistic context, we define the deformation gradient in terms of the radar metric and the mapping from ${\cal S}$ to ${\cal M}$ by
\be\label{deformationgradient}
	F_{\mu i} \equiv f_{\mu\nu} \Xup{\nu}{i}  \ .
\ee
The inverse of the deformation gradient is 
\be
	(F^{-1})^{i \mu} \equiv \Zdown{i}{\nu} g^{\nu\mu}  \ .
\ee
One can check that 
\bse\bea
	F_{\mu i} (F^{-1})^{i\nu} & = & f_\mu^\nu \\
	(F^{-1})^{i\mu} F_{\mu j} & = & \delta^i_j
\eea\ese
using the identities (\ref{orthorelations}). 
The radar metric in matter space, $f_{ij}$, is called the {\em right Cauchy--Green deformation tensor} in continuum mechanics. 
We can write this in terms of the deformation gradient as 
\be
	f_{ij} = F_{\mu i} g^{\mu\nu} F_{\nu j} 
\ee
Using the definitions (\ref{deformationgradient}) for the deformation gradient and (\ref{radarmetric}) for the radar metric, we find the previous 
definition (\ref{radarmetriconS}a) of the radar metric, $f_{ij} = \Xup{\mu}{i} f_{\mu\nu} \Xup{\nu}{j}$.

The {\em Lagrangian strain tensor} (sometimes called the {\em Green strain tensor}) is 
\be\label{Lagrangianstrain}
	E_{ij} = \frac{1}{2} \left( f_{ij} - \epsilon_{ij} \right)
\ee
where the {\em relaxed metric} $\epsilon_{ij}$ is the metric on matter space ${\cal S}$ that characterizes the undeformed body.  That is, 
when the body is relaxed (in flat spacetime with no vibrations, no rotations, and no bulk forces) the proper 
distance $ds$ between neighboring material particles is $ds^2 = \epsilon_{ij} d\zeta^i d\zeta^j$. 

A number of other tensors appear in the literature on continuum mechanics. These tensors are not used in this paper, but 
we present them  here for the sake of completeness. 

The relativistic version of the {\em left Cauchy--Green deformation tensor} (sometimes called the {\em Finger deformation tensor})  is
the spatial tensor 
\be
	B_{\mu\nu} = F_{\mu i} \epsilon^{ij} F_{\nu j} = f_{\mu\alpha}  \Xup{\alpha}{i}  \epsilon^{ij}
	f_{\nu\beta} \Xup{\beta}{j} \ ,
\ee
where $\epsilon^{ij}$ is the inverse of $\epsilon_{ij}$. The inverse of the left Cauchy--Green deformation tensor is 
\be
	(B^{-1})^{\mu\nu} = g^{\mu\sigma} Z^i_{,\sigma} \epsilon_{ij} Z^j_{,\rho} g^{\rho\nu} \ ,
\ee
so that 
\be
	B_{\mu\sigma} (B^{-1})^{\sigma\nu} = f_\mu^\nu \ .
\ee
The {\em Eulerian strain tensor} (also called the {\em Almansi strain tensor}) is
\be
	e_{\mu\nu} = \frac{1}{2} \left( f_{\mu\nu} - \Zdown{i}{\mu} 
	\epsilon_{ij} \Zdown{j}{\nu}  \right) \ .
\ee
Note that the Lagrangian and Eulerian strain tensors satisfy
\bse\bea
	E_{ij} & = & X^\mu_{,i} e_{\mu\nu} X^\nu_{,j} \ ,\\
	e_{\mu\nu} & = & Z^i_{,\mu} E_{ij}  Z^j_{,\nu} \ .
\eea\ese
Thus, the Lagrangian strain is just the Eulerian strain (which is a spatial tensor) mapped to  matter space ${\cal S}$. 

\section{Action and Equations of Motion}\label{actionandeoms}
A material whose  behavior is only a function of the current state of deformation is called {\em elastic}.  If the work  done  by  
stresses  during  the  deformation  process depends only on the initial and final configurations, the material is {\em hyperelastic}. 
In this case we can  introduce an energy density as a function of the Lagrangian strain  $E_{ij}$. 

We will define the energy density $\rho(E)$ as the energy of a deformed material element per unit of physical volume occupied by the 
undeformed (relaxed) element. When the body is relaxed, the physical volume occupied  by the coordinate cell $d^3\zeta$ is defined 
by $\sqrt{\epsilon} d^3\zeta$, where $\epsilon$ is the determinant of the relaxed metric $\epsilon_{ij}$. Thus, 
for the deformed body,  the energy contained in the coordinate cell $d^3\zeta$  is given by $\sqrt{\epsilon} \rho(E) \, d^3\zeta$. 

The relativistic action for a hyperelastic material is \cite{DeWitt62,Brown96}
\be\label{matteraction}
	S[X,g] = -\int_{\lambda_1}^{\lambda_2} d\lambda \int_{\cal S} d^3\zeta \, \sqrt{\epsilon}\, \alpha \rho(E) \ .
\ee
Here, $\alpha$ is the material lapse from Eq.~(\ref{alphadef}), and $E_{ij}$ is the Lagrangian strain. This action defines the system from a Lagrangian perspective, 
That is, the dynamics are described using $x^\mu = X^\mu(\lambda,\zeta)$,  with the matter space coordinates $\zeta^i$ as independent variables and the spacetime 
coordinates $x^\mu$ as dependent variables. 
Recent  work on relativistic elasticity has focused on the Eulerian perspective. In that case the dynamics are described by $\zeta^i = Z^i(x)$, 
with the spacetime coordinates $x^\mu$ as independent variables and the matter space coordinates $\zeta^i$ as dependent variables. 

The energy density $\rho(E)$ will typically depend on the relaxed metric $\epsilon_{ij}$ as well as $E_{ij}$. If the material properties are not uniform, the density 
will depend on the matter space coordinates $\zeta^i$ as well.  The energy density can also depend on other tensors  in matter space, in addition to $E_{ij}$ 
and $\epsilon_{ij}$.  For example, if the body has a crystal lattice structure, then the energy density will depend on a preferred frame (or vector fields) in ${\cal S}$. 
The energy density might also depend on the specific entropy of the material, which would appear as a scalar field in matter space. 
For notational simplicity, we won't normally display the dependence of $\rho(E)$ on $\zeta^i$, $\epsilon_{ij}$, or any other matter space tensors. 

The action  (\ref{matteraction}) depends on the dynamical variables $X^\mu$ directly, and also indirectly through the  argument of the spacetime metric. 
Explicitly, the material lapse $\alpha$, as it appears in the action, is
\be\label{alphadefwithXs}
	\alpha = \sqrt{ - \dot X^\mu \dot X^\nu g_{\mu\nu}(X)} \ .
\ee
The Lagrangian strain $E_{ij}$ depends on the radar metric, which is explicitly given by
\be
	f_{ij} = \Xup{\mu}{i}  \left( g_{\mu\nu}(X) + \frac{1}{\alpha^2}\dot X^\rho \dot X^\sigma g_{\rho\mu}(X) g_{\sigma\nu}(X) \right) \Xup{\nu}{j} \ . 
\ee
The specific properties of the hyperelastic material are determined by the 
functional form of the energy density $\rho(E)$, including its possible dependence on non--dynamical matter space tensors (such as $\epsilon_{ij}$) and 
coordinates $\zeta^i$. 

The action (\ref{matteraction}) is the natural extension of the action for a continuum of non--interacting (dust) particles. In that case 
the energy density $\rho(E)$ is a constant.  We can specialize $S[X]$ to the action for a single relativistic particle
by setting the density to $\rho = (m/\sqrt{\epsilon} ) \delta^3(\zeta - \zeta_0)$, with $\zeta_0^i$ some fixed point in ${\cal S}$. 
Then Eq.~(\ref{matteraction}) reduces to 
\be\label{particleaction}
	S_{particle}[X] = -m\int_{\lambda_1}^{\lambda_2} d\lambda\,  \alpha  
\ee
with $\alpha$ defined in Eq.~(\ref{alphadefwithXs}). The particle action is a functional of $X^\mu(\lambda,\zeta_0)$. 

The equations of motion for the hyperelastic body follow from variation of $S[X,g]$ with respect to the fields $X^\mu(\lambda,\zeta)$. For this calculation, we need 
the results
\bse\label{deltaresults}\bea
	\delta\alpha & = & -U_\mu \delta\dot X^\mu - \frac{1}{2} \alpha U^\mu U^\nu \delta g_{\mu\nu} \ ,\\
	\delta f_{ij}  & = & \frac{2}{\alpha} F_{\mu(i} v_{j)} \delta \dot X^\mu + 2F_{\mu(i} \delta \Xup{\mu}{j)} \nono\\
	& & +  F^\mu_i F^\nu_j \delta g_{\mu\nu}  \ .
\eea\ese
Because the metric is evaluated at the spacetime event $x^\mu = X^\mu(\lambda,\zeta)$, it's variation includes a contribution from the variation of  
the tensor components $g_{\mu\nu}(x)$ as well as a contribution from the variation of its argument: 
\bea
	\delta g_{\mu\nu} & = & \delta g_{\mu\nu}(x) \bigr|_{x = X(\lambda,\zeta)} \nono\\
	& & + \partial_\sigma g_{\mu\nu}(x)\bigr|_{x = X(\lambda,\zeta)} \delta X^\sigma(\lambda,\zeta)  \ .
\eea
The partial derivative of the metric can be written in terms of Christoffel symbols using $\partial_\sigma g_{\mu\nu} = 2\Gamma_{(\mu\nu)\sigma}$. 

The variation of the action (\ref{matteraction}) is
\be
	\delta S = -\int_{\lambda_1}^{\lambda_2} d\lambda \int_{\cal S} d^3\zeta \, \sqrt{\epsilon} \left[ \rho\, \delta\alpha + \frac{\alpha}{2} S^{ij} \delta f_{ij} \right] \ ,
\ee
where 
\be
	S^{ij} \equiv \partial \rho/\partial E_{ij}
\ee
is the {\em second Piola--Kirchhoff stress tensor}. (Stress is discussed in more detail in the next section.) 
Using the results for $\delta\alpha$, $\delta f_{ij}$, and $\delta g_{\mu\nu}$ from above, we find 
\bea
	\delta S & = & \int_{\lambda_1}^{\lambda_2} d\lambda \int_{\cal S} d^3\zeta \, \sqrt{\epsilon} \biggl[ ( \rho U_\mu - S^{ij} F_{\mu i}v_j ) \delta \dot X^\mu \nono\\
	& & \quad - \alpha S^{ij} F_{\mu i} \delta\Xup{\mu}{j}  \nono\\
	& & \quad + \alpha ( \rho U^\alpha U^\beta - S^{ij} F^\alpha_i F^\beta_j ) \Gamma_{\alpha\beta\mu} \delta X^\mu \biggr]  \ .
\eea
where $F_{\mu i}$ is the deformation gradient (\ref{deformationgradient}). The next step in deriving the equations of motion 
is to remove the derivatives 
from $\delta X^\mu$ through integration by parts. This generates endpoint terms in $\delta S$ at the initial and final parameter values $\lambda_1$ and $\lambda_2$,  as well as terms on the boundary of matter space $\partial{\cal S}$. 
With the initial and final configurations of the elastic body fixed,  the variations in $X^\mu$ vanish at $\lambda_1$ and $\lambda_2$. 
This ensures that the endpoint terms in  $\delta S $ vanish. For the matter space boundary $\partial{\cal S}$, we assume that the surface of the elastic body is free. 
These are referred to as ``natural" boundary conditions \cite{Lanczos} since they arise naturally from the variational principle. In the language of continuum mechanics, these are called  force/stress or traction boundary conditions, with the external force chosen to vanish.   Thus, the variation of the action becomes
\bea
	\delta S & = & \int_{\lambda_1}^{\lambda_2} d\lambda \int_{\cal S} d^3\zeta \biggl[ -\sqrt{\epsilon} \frac{\partial}{\partial\lambda} 
	( \rho U_\mu - S^{ij} F_{\mu i} v_j)  \nono\\
	& & \quad + \frac{\partial}{\partial\zeta^j} ( \sqrt{\epsilon}\alpha S^{ij} F_{\mu i} )  \nono\\
	& & \quad + \sqrt{\epsilon}\alpha (\rho U^\alpha U^\beta - S^{ij} F^\alpha_i F^\beta_j) \Gamma_{\alpha\beta\mu} \biggr] \delta X^\mu \nono\\
	& & -  \int_{\lambda_1}^{\lambda_2} d\lambda  \int_{\partial{\cal S}} d^2\eta \sqrt{\theta} \, \alpha\, n_j S^{ij} F_{\mu i} \delta X^\mu \ .
\eea
Here, $\eta^A$ (with $A = 1,2$) are coordinates on the matter space boundary $\partial{\cal S}$. The metric on the boundary has determinant $\theta$, and the 
unit normal to the boundary is $n_i$. These are defined using the relaxed metric $\epsilon_{ij}$. 

Setting the variation of the action to zero, the volume integral term gives the ``bulk" equation of motion 
\be\label{bulkeoms}
	0 = -\sqrt{\epsilon} \frac{D}{D\lambda} ( \rho U_\mu - S^{ij} F_{\mu i} v_j) 
	+ \frac{D}{D\zeta^j} ( \sqrt{\epsilon}\alpha S^{ij} F_{\mu i} )
\ee
where the covariant derivatives with respect to $\lambda$ and $\zeta^i$ are defined by\footnote{These definitions use a common 
abuse of notation. Consider a tensor $T(\lambda,\zeta)$ with spacetime indices suppressed. The right--hand side of Eq.~(\ref{Ddefs}a) is shorthand notation 
for $\dot X^\mu( \nabla_\mu \tilde T(x) )\bigr|_{x = X(\lambda,\zeta)}$, where  $\tilde T$ is defined by 
$\tilde T(x) = T(\Lambda(x),Z(x))$. A similar definition holds for the right--hand side of Eq.~(\ref{Ddefs}b).}
\bse\label{Ddefs}\bea
	\frac{D}{D\lambda} & = & \dot X^\mu \nabla_\mu \ ,\\
	 \frac{D}{D\zeta^i}  & = & \Xup{\mu}{i} \nabla_\mu \ .
\eea\ese
With $\delta S = 0$, the integral over the matter space boundary  gives
\be
	0 = -n_j S^{ij} F_{\mu i} \bigr|_{\partial{\cal S}} \ .
\ee
These are the natural boundary conditions. 

We can write these equations is slightly more compact form by introducing the {\em first Piola--Kirchhoff stress}:
\be\label{firstPKstress}
	P_{\mu}^i \equiv F_{\mu j} S^{ij} \ .
\ee
Then the bulk equations of motion and natural boundary conditions become
\bse\label{equationsofmotion}\bea
	\sqrt{\epsilon} \frac{D}{D\lambda} ( \rho U_\mu - P^j_\mu v_j) 
	& = & \frac{D}{D\zeta^j} ( \sqrt{\epsilon}\alpha P^j_\mu ) \ ,\\
	P^i_\mu n_i \bigr|_{\partial{\cal S}} & = & 0 \ .
\eea\ese

\section{Stress, Energy and Momentum}\label{SEMsection}
The stress--energy--momentum (SEM) tensor for matter (non--gravitational) fields is defined by 
\be
	T^{\mu\nu}(x) = \frac{2}{\sqrt{-g}} \frac{\delta S_{matter}}{\delta g_{\mu\nu}(x)} \ .
\ee
The functional derivative of the matter action $S_{matter}$ is determined by the coefficient of $\delta g_{\mu\nu}(x)$ in the variation $\delta S_{matter}$. 
To apply this definition to the elastic material, we first write the action as an integral over the spacetime coordinates: 
\be
	S[X,g] =  -\int d^4x \int_{\lambda_1}^{\lambda_2} d\lambda \int_{\cal S} d^3\zeta \, \sqrt{\epsilon} \alpha \rho \, \delta^4(x - X(\lambda,\zeta)) \ .
\ee
The Dirac delta function enforces the evaluation of the metric $g_{\mu\nu}(x)$ at the spacetime event $X^\mu(\lambda,\zeta)$. Using the results for $\delta\alpha$ 
and $\delta f_{ij}$ from  Eqs.~(\ref{deltaresults}), we find 
\bea\label{semwithdelta}
	T^{\mu\nu}(x) & = & \int_{\lambda_i}^{\lambda_f} d\lambda \int_{\cal S} d^3\zeta \frac{\alpha\sqrt{\epsilon}}{\sqrt{-g}} \delta^4(x - X(\lambda,\zeta)) \nono\\
	& & 	\qquad\qquad\quad \left[ \rho U^\mu U^\nu - S^{ij} F^\mu_i F^\nu_j \right] \ .
\eea
We can evaluate the stress--energy--momentum tensor at the event $x^\mu = X^\mu(\bar\lambda,\bar\zeta)$. The Dirac delta function becomes 
\be\label{Diracdeltaresult}
	\delta^4(X(\bar\lambda,\bar\zeta) - X(\lambda,\zeta)) = \frac{1}{|\det(\Xup{\cdot}{\cdot})|} \delta(\bar\lambda - \lambda) \delta^3(\bar\zeta - \zeta) \ ,
\ee
where $\det(\Xup{\cdot}{\cdot})$ is the determinant of the matrix formed from derivatives of $X^\mu$ with respect to $\lambda$ and $\zeta^i$. 
Then the SEM tensor becomes 
\be\label{semtensor}
	T^{\mu\nu}(X(\bar\lambda,\bar\zeta)) = \frac{\alpha\sqrt{\epsilon}}{\sqrt{-g}   |\det(\Xup{\cdot}{\cdot})|   } \left[ \rho U^\mu U^\nu - S^{ij} F^\mu_i F^\nu_j \right] 
\ee
where the right--hand side is evaluated at $\bar\lambda$ and $\bar\zeta^i$. We can, of course, drop the bars and rewrite this result by setting 
$T^{\mu\nu}(X(\lambda,\zeta))$ equal to the right--hand side above, with the right--hand side now evaluated at $\lambda$, $\zeta^i$.

The factor $|\det(\Xup{\cdot}{\cdot})|$ can be analyzed by considering the worldline parameter $\lambda$ and the matter space coordinates $\zeta^i$, together, 
to define a coordinate system on spacetime ${\cal M}$ within the world tube of the body. Denote this coordinate system by $x^{\mu'} = \{ \lambda,\zeta^i \}$, so the 
mapping from $\Re \times {\cal S}$ to ${\cal M}$ defines a coordinate transformation $x^\mu = X^\mu(x')$. 
The components of the metric in the primed coordinates are 
\be
	g'_{\mu\nu} = \frac{\partial X^\alpha}{\partial x^{\mu'}} \frac{\partial X^\beta}{\partial x^{\nu'}} g_{\alpha\beta} \ .
\ee
Taking the determinant of this relation we find 
\be\label{detXandgprime}
	|\det(\Xup{\cdot}{\cdot})| = \sqrt{-g'}/\sqrt{-g} \ .
\ee
Now we can use the definitions from Sec.~(\ref{kinematics}) for $\alpha$, $v_i$ and $f_{ij}$ to compute 
\be
	g'_{\mu\nu} = \begin{pmatrix} -\alpha^2 & \alpha v_i \\ \alpha v_j & f_{ij} - v_i v_j \end{pmatrix}
\ee
The determinant of $g'_{\mu\nu}$ follows from the formula 
\be
	\det \begin{pmatrix} A & B \\ C & D \end{pmatrix} = \det(A) \, \det(D - C A^{-1} B )
\ee
for the determinant of a block matrix. This yields 
\be
	\det(g') = -\alpha^2 f
\ee
where $f$ is the determinant of the radar metric $f_{ij}$. Putting this together with Eq.~(\ref{detXandgprime}) gives  the result 
\be\label{detXdotdot}
	|\det(\Xup{\cdot}{\cdot})| = \alpha\sqrt{f}/\sqrt{-g} \ .
\ee 
Then the SEM tensor (\ref{semtensor}) becomes 
\be\label{semtensorfinal}
	T^{\mu\nu}(X(\lambda,\zeta)) = \frac{1}{J} \left[ \rho U^\mu U^\nu - S^{ij} F_i^\mu F_j^\nu \right] \ ,
\ee
where we have defined 
\be\label{Jdefinition}
	J \equiv \sqrt{f}/\sqrt{\epsilon} \ .
\ee
Recall that $\sqrt{\epsilon} d^3\zeta$ is the physical volume occupied by the coordinate cell $d^3\zeta$ when the body is in its relaxed state. Similarly, 
$\sqrt{f} d^3\zeta$ is the physical volume occupied by the coordinate cell $d^3\zeta$ when the body is deformed. Thus, the factor $1/J$ 
in Eq.~(\ref{semtensorfinal}) converts the energy per unit relaxed volume (the dimensions of $\rho$ and $S^{ij}$) into the energy per unit 
deformed volume (the dimensions of $T^{\mu\nu}$). 

Consider a comoving observer, that is, an observer whose worldline coincides with a particular particle $\zeta^i$ in the body.  The observer's 
velocity is $U^\mu$ and their spatial directions are spanned by $\Zup{i}{\mu}$. The energy density as seen by this observer is 
\be
	T^{\mu\nu} U_\mu U_\nu =  \rho/J \ .
\ee
The energy flux (momentum density) for this observer vanishes:  $T^{\mu\nu} U_\mu \Zdown{i}{\nu} = 0$. 
The momentum flux (spatial stress) is 
\be
	T^{\mu\nu} \Zdown{i}{\mu} \Zdown{j}{\nu} = -  S^{ij} /J \ .
\ee
Note the relative minus sign between the spatial components of the SEM tensor and the second Piola--Kirchhoff stress tensor. 
This arises because the stress components of $T^{\mu\nu}$  give the $i$--component of force that the material on the $\zeta^j < {\rm const}$ side 
of the surface $\zeta^j = {\rm const}$ exerts on the $\zeta^j > {\rm const}$ side. The second Piola--Kirchhoff stress is defined with the opposite convention, 
as the force that the $\zeta^j > {\rm const}$ side exerts on the $\zeta^j < {\rm const}$ side. 

Of course the elastic body stress--energy--momentum tensor must satisfy the local conservation law $\nabla_\mu T^{\mu\nu} = 0$. We can verify this by explicit 
calculation using Eq.~(\ref{semwithdelta}). First, expand the covariant derivative as 
\be\label{nablaT}
	\nabla_\mu T^{\mu\nu} = \frac{1}{\sqrt{-g}} \partial_\mu (\sqrt{-g} T^{\mu\nu}) + \Gamma^\nu{}_{\mu\alpha} T^{\mu\alpha} \ .
\ee
The partial derivative with respect to $x^\mu$ acts on the Dirac delta function in Eq.~(\ref{semwithdelta}) to give 
$\partial_\mu \delta^4(x - X(\lambda,\zeta))$. The index
 $\mu$ is contracted with either $U^\mu$ or $F^\mu_i$. Recall that the spacetime velocity $U^\mu$ is proportional to $\dot X^\mu$. 
 Using the definitions of Sec.~\ref{kinematics},  the deformation gradient (\ref{deformationgradient}) can be rewritten  as 
\be
	F^\mu_i = f^\mu_\alpha \Xup{\alpha}{i} = \Xup{\mu}{i} +\frac{1}{\alpha}  v_i \dot X^\mu  \ .
\ee
Thus, $F^\mu_i$ is a combination of terms that are proportional to $\dot X^\mu$ and $\Xup{\mu}{i}$. So the $\mu$ index in
$\partial_\mu \delta^4(x - X(\lambda,\zeta))$ is always contracted with either $\dot X^\mu$ or $\Xup{\mu}{i}$. 
We can rewrite these expressions using the following identities:
\begin{widetext}
\bse\label{derivativeXDiracdelta}\bea
 	\dot X^\mu  \partial_\mu \delta^4(x - X(\lambda,\zeta)) & = &  \dot X^\mu \left[ - \frac{\partial}{\partial X^\mu} \delta^4 (x - X) \right] \biggr|_{X = X(\lambda,\zeta)}  
	= -\frac{\partial}{\partial\lambda} \delta^4(x - X(\lambda,\zeta))  \ ,\\
	\Xup{\mu}{i} \partial_\mu \delta^4(x - X(\lambda,\zeta)) & = & \Xup{\mu}{i} \left[ - \frac{\partial}{\partial X^\mu} \delta^4 (x - X) \right] \biggr|_{X = X(\lambda,\zeta)}  
	= -\frac{\partial}{\partial\zeta^i} \delta^4(x - X(\lambda,\zeta)) \ .
\eea\ese
The next step in evaluating  $\nabla_\mu T^{\mu\nu}$ is to integrate by parts to shift the derivatives $\partial/\partial\lambda$ and $\partial/\partial\zeta^i$ 
away from the Dirac delta function. The endpoint and boundary terms can be discarded if we choose the spacetime point $x^\mu$ inside the world tube 
of the body. The result that follows from Eq.~(\ref{nablaT})  is
\be
	\nabla_\mu T^{\mu\nu}(x) = \frac{1}{\sqrt{-g}} \int_{\lambda_i}^{\lambda_f} d\lambda \int_{\cal S} d^3\zeta \left\{\sqrt{\epsilon} \frac{D}{D\lambda} ( \rho U^\nu 
	- S^{ij} F^\nu_i v_j ) - \frac{D}{D\zeta^i} ( \sqrt{\epsilon}  \alpha S^{ij} F^\nu_j ) \right\} \delta^4(x - X(\lambda,\zeta)) \ .
\ee
We can evaluate this expression at the point $X^\mu(\bar\lambda,\bar\zeta)$ inside the world tube of the body, then rewrite the Dirac delta function as in 
Eqs.~(\ref{Diracdeltaresult}) and (\ref{detXdotdot}).   Carrying out the integrals over $\lambda $ and $\zeta^i$, then dropping the bars on $\bar\lambda$ 
and $\bar \zeta^i$, we have 
\be
	\nabla_\mu T^{\mu\nu} (X(\lambda,\zeta)) = \frac{1}{\alpha\sqrt{f}} \left\{\sqrt{\epsilon} \frac{D}{D\lambda} ( \rho U^\nu 
	- S^{ij} F^\nu_i v_j ) - \frac{D}{D\zeta^i} ( \sqrt{\epsilon}  \alpha S^{ij} F^\nu_j ) \right\} \ .
\ee
\end{widetext}
The term in curly brackets vanishes when the bulk equations of motion  (\ref{bulkeoms}) are satisfied, so the equations of motion 
insure that the local conservation law $\nabla_\mu T^{\mu\nu} = 0$ holds. More than that, we see that the bulk equations of motion for a 
hyperelastic material coincide  with the conservation law $\nabla_\mu T^{\mu\nu} = 0$. 

\section{Point Particle}\label{ppsection}
It will be useful to review the simple example of a relativistic point particle. If we choose the energy density as $ \rho = (m/\sqrt{\epsilon}) \delta^3(\zeta - \zeta_0)$, 
then the elastic body action (\ref{matteraction}) reduces to the action (\ref{particleaction}) for a 
single particle of mass $m$ (located at the point $\zeta_0^i$ in matter space).  

The particle action is invariant under reparametrizations of the worldline \cite{HenneauxTeitelboim}. This is a type of gauge symmetry, which can be understood as follows. 
Consider worldline parameters $\lambda_1$ and $\lambda_2$ related by $\lambda_1 = \Lambda(\lambda_2)$.\footnote{Do not confuse 
$\Lambda(\lambda)$  with the function $\Lambda(x)$ from Sec.~(\ref{kinematics}).} 
Given a history $X^\mu(\lambda)$,  define a new history $\tilde X^\mu(\lambda)$ by 
\be\label{particlereparaminv}
	\tilde X^\mu(\lambda) = X^\mu(\Lambda(\lambda))
\ee
Derivatives of these histories are related by 
\be
	\frac{\partial\tilde X(\lambda)}{\partial\lambda} = \frac{\partial X(\lambda)}{\partial\lambda}\biggr|_{\lambda = \Lambda(\lambda)} 
	\frac{\partial\Lambda(\lambda)}{\partial\lambda}
\ee
From this result, we find that the action for the history $\tilde X^\mu(\lambda)$ is 
\be
	S_{particle}[\tilde X] = -m \int_{\lambda_i}^{\lambda_f} d\lambda  \, \alpha \bigr|_{\lambda = \Lambda(\lambda)}
\ee
If the reparametrization is the identity at the endpoints, so that $\Lambda(\lambda_i) = \lambda_i$ and $\Lambda(\lambda_f) = \lambda_f$, then a simple 
change of integration variables from $\lambda$ to $\Lambda$ shows that $S[\tilde X] = S[X]$. That is, the action is the same for any two histories that are 
related by the reparametrization (\ref{particlereparaminv}) of the worldline.  The action is gauge invariant. 

For the single point particle, the 
equations of motion (\ref{equationsofmotion}) reduce to the geodesic equations, 
$D U^\mu/D\lambda  = 0$, where $U^\mu \equiv \dot X^\mu/\alpha$. Using the identity 
$\alpha\, DU^\mu/D\lambda = (\delta^\mu_\nu + U^\mu U_\nu) D\dot X^\nu/D\lambda$, the equations of motion become 
\be\label{particleeoms}
	 (\delta^\mu_\nu + U^\mu U_\nu) \frac{D}{D\lambda} \dot X^\nu = 0 \ .
\ee
Although there are four equations of motion, only three are independent. In particular, the linear combination obtained by contracting 
Eqs.~(\ref{particleeoms}) with $U_\mu$ vanishes. Said another way, the equations of motion cannot be solved for all of the $\ddot X^\mu$'s 
as functions of $X^\mu$ and $\dot X^\mu$. Given initial data $X^\mu(0)$ and $\dot X^\mu(0)$, the future evolution is 
not fully determined because the worldline parameter is arbitrary. 

One way to choose the parametrization (to ``fix the gauge") is to set $\lambda$ equal to proper time. Then  $\alpha = 1$ and  $\dot X^\mu$ equals the 
spacetime velocity $U^\mu$. Note that in this gauge, $\dot X^\mu$ is normalized: $\dot X^\mu g_{\mu\nu} \dot X^\nu = -1$. The four second derivatives, $\ddot X^\mu$, 
are now determined by the three independent equations (\ref{particleeoms}) plus the covariant $\lambda$--derivative of the normalization 
condition: 
\be\label{derivnormcond}
	\dot X_\mu \frac{D}{D\lambda} \dot X^\mu = 0 \ .
\ee
Together, these equations yield the geodesic equations $D\dot X^\mu/D\lambda = 0$ with affine 
parametrization.\footnote{We are
not allowed to set the parameter $\lambda$ equal to proper time in the action because this would fix the proper time interval between initial and final configurations.  
The variational principle must allow for histories with differing proper time intervals.}

The worldline parameter can be chosen in a convenient way if one of the spacetime coordinates, say $t \equiv x^0$, has the property 
that the $t = {\rm const}$ surfaces are spacelike. That is, the coordinate basis vectors $\partial/\partial x^a$ with 
$a = 1,2,3$ are spacelike. (Note that the coordinate basis vector $\partial/\partial t$ need not be timelike.) In this case, 
we can choose $X^0(\lambda) = \lambda$. It follows that 
$\dot X^0 = 1$. In this gauge the dynamical variables are the spatial components $X^a(t)$  of the particle worldline, where $a = 1,2,3$. 
	
The action in this gauge is most conveniently expressed using the ADM decomposition of the spacetime metric:
\be\label{ADMdecomp}
	g_{\mu\nu} = \begin{pmatrix} N_a N^a - N^2 & N_a \\ N_b & g_{ab} \end{pmatrix} \ ,
\ee
where $N$ is the spacetime lapse function (not to be confused with the material lapse function $\alpha$) and $N^a$ is the shift vector. 
Indices on $N^a$ are raised and lowered with the spatial metric $g_{ab}$.  
In the $\lambda = t$ gauge the material lapse (\ref{alphadef}) becomes
\be\label{alphatgauge}
	\alpha  = \left[ N^2 - (\dot X^a + N^a) g_{ab} (\dot X^b + N^b) \right]^{1/2} \ . 
\ee
This can be written more simply by defining 
\be\label{Vdefinition}
	V^a \equiv (\dot X^a + N^a)/N  \ ,
\ee
which are the spatial components  (in the coordinate basis $\partial /\partial x^a$) of the particle velocity as seen by observers 
at rest in the $t = {\rm const}$ surfaces. Then $\alpha = N\sqrt{1 - V^a V_a}$  and we see that 
\be\label{gammafactor}
	\gamma \equiv N/\alpha = 1/\sqrt{1 - V^a V_a}
\ee
is the relativistic gamma factor  between the particle and the observers at rest in $t = {\rm const}$ surfaces. Note that indices on $V^a$ are lowered with 
the spatial metric $g_{ab}$. 

The action (\ref{matteraction}) in the $\lambda = t$  gauge reduces to 
\be\label{particleactiongaugefixed}
	S_{particle}[X] = -m\int_{t_i}^{t_f} dt  \,  N\sqrt{1 - V^a V_a}   \ . 
\ee
The spacetime metric components $N$, $N^a$ and $g_{ab}$, as they appear in the action, are functions of $t$ and $X^a(t)$. 
The action is a functional  of $X^a(t)$. 

In the gauge $\lambda = t$, the equations of motion are
\be\label{particleeomsgauge}
	D_t (\gamma V_a) + \gamma \partial_a N - \gamma V_b D_a N^b = 0 \ ,
\ee
where $D_t$  and $D_a$ are covariant derivatives compatible with the spatial metric $g_{ab}$. Explicitly, we have 
$D_t(\gamma V_a) = \partial_t (\gamma V_a) -  {}^{(3)}\Gamma_{ab}^c  (\gamma V_c) \dot X^b$ where ${}^{(3)}\Gamma_{ab}^c$ are the 
Christoffel symbols constructed from $g_{ab}$.  
The result (\ref{particleeomsgauge}) 
is most easily obtained by extremizing the action (\ref{particleactiongaugefixed}). Alternatively, we can set  $\lambda = t$ in the gauge invariant equations 
(\ref{particleeoms}) and  make use of the results
\bse\label{UupandUdown}\bea
	U^a & = & \dot X^a/\alpha \ , \quad U_a = \gamma V_a \ ,\\
	U^0 & = & 1/\alpha \ , \quad U_0 = \gamma(N_a V^a - N) \ ,
\eea\ese
for the covariant and contravariant components of the spacetime velocity. 

\section{Gauge Fixed Theory}\label{gaugesection}
The formal structure of the relativistic elastic theory is closely analogous to that of the relativistic particle. Let two worldline parameters  $\lambda_1$ and $\lambda_2$
be related by 
\be\label{reparaminv}
	\lambda_1 = \Lambda(\lambda_2,\zeta) \ .
\ee
Note that $\Lambda$ depends on $\zeta^i$; the parameter for each particle in the body can be changed independently from one another, restricted only by 
continuity. Define a new history $\tilde X^\mu(\lambda,\zeta)$ related to the old history $X^\mu(\lambda,\zeta)$ by
\be	
	\tilde X^\mu(\lambda,\zeta) = X^\mu(\Lambda(\lambda,\zeta),\zeta) \ .
\ee
Derivatives of  these histories are related by 
\bse\bea
	\frac{\partial\tilde X^\mu(\lambda,\zeta)}{\partial \lambda} & = & \frac{\partial X^\mu(\lambda,\zeta)}{\partial \lambda} \biggr|_{\lambda = \Lambda} 
	\frac{\partial\Lambda}{\partial\lambda}  \ ,\\
	\frac{\partial\tilde X^\mu(\lambda,\zeta)}{\partial\zeta^i} & = & \frac{\partial X^\mu(\lambda,\zeta)}{\partial\lambda} \biggr|_{\lambda = \Lambda} 
	\frac{\partial\Lambda}{\partial\zeta^i} \nono\\
	& &  + \frac{\partial X^\mu(\lambda,\zeta)}{\partial \zeta^i} \biggr|_{\lambda = \Lambda} 
\eea\ese
with $\Lambda \equiv \Lambda(\lambda,\zeta)$. From these results we find that the action for $\tilde X^\mu(\lambda,\zeta)$ is 
\be
	S[\tilde X]  = -\int_{\lambda_i}^{\lambda_f} d\lambda \int_{\cal S} d^3\zeta \, \frac{\partial\Lambda(\lambda,\zeta)}{\partial\lambda} 
	\sqrt{\epsilon} \, \alpha \rho \bigr|_{\lambda = \Lambda(\lambda,\zeta)} \ .
\ee
Let the reparametrization become the identity at the endpoints: $\Lambda(\lambda_i,\zeta) = \lambda_i$ and $\Lambda(\lambda_f,\zeta) = \lambda_f$, Then a 
simple change of integration variables, with
$d\Lambda\, d^3\zeta = d\lambda \,d^3\zeta\, (\partial\Lambda/\partial\lambda)$, shows that $S[\tilde X] = S[X]$. Thus the action for the elastic body 
is the same for any two histories that are related by the reparametrization (\ref{reparaminv}) of the worldlines. The action is gauge invariant. 

The equations of motion that follow from the gauge invariant action are listed in Eqs.~(\ref{equationsofmotion}). We want to show that the linear combination 
of bulk equations (\ref{equationsofmotion}a)  obtained by contracting with $U^\mu$ is vacuous; that is, simply $0 = 0$. Begin by contracting both sides 
with $U^\mu$ and  use the fact that $D/D\lambda$ and $D/D\zeta^j$ obey the product rule of differentiation. Since $U^\mu P^j_\mu = 0$, we find (after dropping 
an overall factor of $-\sqrt{\epsilon}$)
\be\label{showthisisvacuous}
	 \frac{D\rho}{D\lambda} + \frac{DU^\mu}{D\lambda} (\rho U_\mu - P^j_\mu v_j) 
	= \frac{DU^\mu}{D\zeta^j} \alpha P^j_\mu \ .
\ee
From the definitions $S^{ij} = \partial\rho/\partial E_{ij}$ and $E_{ij} = (f_{ij} - \epsilon_{ij})/2$ for the second Piola--Kirchhoff stress and the Lagrangian strain, 
the first term above becomes
\be
	\frac{D\rho}{D\lambda} = \frac{1}{2} S^{ij} \frac{\partial f_{ij}}{\partial \lambda} \ .
\ee
The spacetime velocity is defined by $U^\mu = \dot X^\mu/\alpha$ with $\alpha = \sqrt{-\dot X^\mu \dot X_\mu}$; from this we find 
\bse\label{DUresults}\bea
	\frac{DU^\mu}{D\lambda} & = & \frac{1}{\alpha} f^\mu_\nu  \frac{D \dot X^\nu}{D\lambda} \ ,\\
	\frac{DU^\mu}{D\zeta^j} & = & \frac{1}{\alpha} f^\mu_\nu  \frac{D \dot X^\nu}{D\zeta^j} \ . 
\eea\ese
With the result (\ref{DUresults}a), we can compute the derivative of the radar metric (\ref{radarmetriconS}a):  
\be\label{dfdlambda}
	\frac{\partial f_{ij}}{\partial \lambda} = 2 \Xup{\mu}{i} f_{\mu\nu} \frac{D\Xup{\nu}{j}}{D\lambda} 
	+ \frac{2}{\alpha} v_i \Xup{\mu}{j} f_{\mu\nu}  \frac{D\dot X^\nu}{D\lambda} \ .
\ee
Using the results (\ref{DUresults}) and (\ref{dfdlambda}), we find that  Eq.~(\ref{showthisisvacuous}) simplifies to 
\be
	f_{\mu\nu} \Xup{\mu}{i} S^{ij} \frac{D \Xup{\nu}{j} }{D\lambda} = f_{\mu\nu} \Xup{\mu}{i} S^{ij} \frac{D \dot X^\nu}{D\zeta^j}
\ee
It is not difficult to verify that the $\lambda$ and $\zeta^i$ derivatives acting on $X^\nu$ commute: $D\Xup{\nu}{j}/D\lambda = D\dot X^\nu/D\zeta^j$. 
Therefore, the equation of motion (\ref{showthisisvacuous}) is indeed vacuous; it reduces to $0 = 0$. In turn, this tells us that only three of the four elastic body 
equations of motion (\ref{equationsofmotion}a)  are independent. 

The gauge (reparametrization) invariance can be fixed by setting $\lambda$ equal to proper time. Then $\alpha = 1$ and $\dot X^\mu = U^\mu$. The four 
second derivatives $\ddot X^\mu$ are determined by the three independent equations (\ref{equationsofmotion}a) plus the derivative of the normalization 
condition, Eq.~(\ref{derivnormcond}). 

As with the relativistic particle, we can fix the gauge by setting $\lambda = t$  where the coordinate $t \equiv x^0$ has the property that the 
$t = {\rm const}$ surfaces are spacelike. Then  $\dot X^0 = 1$ and $\Xup{0}{i} = 0$. The evolution of the elastic body is described by  
$X^a(t,\zeta)$. 

With the ADM metric splitting (\ref{ADMdecomp}), the material lapse $\alpha$ is given by Eq.~(\ref{alphatgauge}). Using the definition 
\be\label{defofVa}
	V^a \equiv (\dot X^a + N^a)/N \ ,
\ee
we find $\alpha = N\sqrt{1 - V^a V_a}$ and the relativistic gamma factor is $\gamma = N/\alpha$. 
The expressions (\ref{UupandUdown}) for the covariant and contravariant components of the spacetime velocity hold in this case as well. 

In this $\lambda = t$ gauge for the elastic body, we have the following useful results for the matter space velocity $v_i$ and the radar metric $f_{ij}$:
\bse\label{vandfresuts}\bea
	v_i & = & \gamma V_a\Xup{a}{i} \ ,\\
	f_{ij} & = &  \Xup{a}{i} (g_{ab} + \gamma^2 V_a V_b)\Xup{b}{j} \ .
\eea\ese
Note that $f_{ab} =  (g_{ab} + \gamma^2 V_a V_b)$ are the spatial components of $f_{\mu\nu}$. The action in this gauge is 
\be\label{matteractioningauge}
	S[X] = -\int_{t_i}^{t_f} dt \int_{\cal S} d^3\zeta \, \sqrt{\epsilon} N \sqrt{1 - V^a V_a} \rho(E) \ ,
\ee
where $E_{ij} = (f_{ij} - \epsilon_{ij})/2$ is the Lagrangian strain. This action is a functional of $X^a(t,\zeta)$. 

The elastic body equations of motion in this gauge are most easily obtained by extremizing the action. Using the relations above, we find 
\begin{widetext}
\be\label{lambdaequalsteoms}
	\sqrt{\epsilon} D_t ( \gamma\rho V_a - v_i P^i_a) - D_j(\sqrt{\epsilon}\alpha P^j_a) + \sqrt{\epsilon} \gamma ( \rho - S^{ij} v_i v_j) \partial_a N 
	- \sqrt{\epsilon} ( \gamma \rho V_b - v_i P^i_b) D_a N^b = 0
\ee
\end{widetext}
where $D_a$ is the covariant derivative compatible with the spatial metric $g_{ab}$ and 
\bse\bea
	D_t & = & \dot X^a D_a \ ,\\
	D_i & = & \Xup{a}{i} D_a \ .
\eea\ese
We also make use of the definition 
\be
	P^i_a = (g_{ab} + \gamma^2 V_a V_b) \Xup{b}{j} S^{ij} \ ;
\ee
these are the spatial components of the first Piola--Kirchhoff stress (\ref{firstPKstress}). Note that the equations of motion (\ref{lambdaequalsteoms}) 
reduce to the single particle equations (\ref{particleeomsgauge}) when $\sqrt{\epsilon} \rho = m\, \delta^3(\zeta - \zeta_0)$. 

\section{Nonrelativistic Limit}\label{NRsection}
Consider the nonrelativistic limit of the elastic  theory in the $\lambda = t$ gauge, as defined by 
the action (\ref{matteractioningauge}) and equations of motion (\ref{lambdaequalsteoms}). 

Let square brackets denote the dimensions of a quantity, where $L$ is length, $T$ is time and $M$ is mass. For example, the speed of light and  Newton's 
gravitational constant have dimensions $[c] = L/T$ and $[G] = L^3/(MT^2)$. 
We will assume that the spacetime coordinates are $t$ and $x^a$, with dimensions $[t] = T$ and $[x^a] = L$. 
Let the matter space coordinates have dimensions $[\zeta^i] = L$. With these choices, the spatial metric $g_{ab}$, radar metric $f_{ij}$, relaxed metric 
$\epsilon_{ij}$ and Lagrangian strain $E_{ij}$  are all dimensionless. 
The spacetime lapse function $N$ and shift vector $N^a$ are defined by the ADM splitting (\ref{ADMdecomp}), with $N$  replaced by $c N$ and $N^a$ 
replaced  $c N^a$. The factors of $c$ compensate
for the change in the ``time" coordinate from $x^0$ (with dimensions $L$) to $t$ (with dimensions $T$). Then the lapse  $N$ and shift  $N^a$ are dimensionless. 

With the above choices,  $V^a$ and $v_i$ have dimensions of velocity,  $[V^a] = [v_i] = L/T$. The definition (\ref{defofVa}) becomes 
$V^a = (\dot X^a + cN^a)/N$.  The 
relativistic gamma factor is defined by  $\gamma = 1/\sqrt{1 - V^a V_a/c^2}$, and  the factor $V_a V_b$ in the 
result (\ref{vandfresuts}b) for $f_{ij}$ must be divided by $c^2$. 

The energy density has dimensions $[\rho(E)] = M/(LT^2)$.  The second and first Piola--Kirchhoff  stress tensors, $S^{ij}$ and $P^i_\mu$, have 
dimensions $M/(LT^2)$ as well. 

Inserting the appropriate factors of $c$ into the elastic body equations of motion (\ref{lambdaequalsteoms}), we have 
\begin{widetext}
\be\label{eomswithc}
	\frac{1}{c^2}\sqrt{\epsilon} D_t ( \gamma\rho V_a - v_i P^i_a) - D_j(\sqrt{\epsilon}\alpha P^j_a) + \sqrt{\epsilon} \gamma ( \rho - S^{ij} v_i v_j /c^2) \partial_a N 
	- \frac{1}{c}\sqrt{\epsilon} ( \gamma \rho V_b - v_i P^i_b) D_a N^b = 0 \ .
\ee
\end{widetext}
The nonrelativistic limit is obtained by writing the spacetime metric as
\be
	g_{\mu\nu} dx^\mu dx^\nu = -(c^2 + 2\Phi) dt^2 + g_{ab} dx^a dx^b \ ,
\ee
setting the matter density to 
\be\label{NRenergydensity}
	\rho(E) = \rho_0 c^2 + W(E) \ .
\ee
and letting $c \to \infty$. 
Here, $\Phi$ is the Newtonian gravitational potential  (with dimensions $L^2/T^2$) and $g_{ab}$ is the flat spatial metric. 
Also, $\rho_0$ is the rest mass per unit undeformed volume (with dimensions $M/L^3$) and $W(E)$ is the potential energy per unit 
undeformed volume (with dimensions $M/(LT^2)$). 

For the spacetime metric above, the spacetime lapse is 
\be\label{NRlapse}
	N = \sqrt{1 + 2\Phi/c^2}
\ee
and the shift vanishes: $N^a = 0$. Inserting these into Eqs.~(\ref{eomswithc})  
and letting $c\to\infty$, we find 
\be\label{NReoms}
	\sqrt{\epsilon} D_t (\rho_0 \dot X_a) - D_j(\sqrt{\epsilon} P^j_a) + \sqrt{\epsilon} \rho_0 \partial_a \Phi = 0  
\ee
where   $V^a = \dot X^a$. Also note that  in this limit, $\alpha = 1$ and the first Piola--Kirchhoff stress becomes
\be
	P^i_a = g_{ab} \Xup{b}{j}S^{ij} \ .
\ee
The second Piola--Kirchhoff stress $S^{ij} = \partial W(E)/\partial E_{ij}$ is defined in terms of the Lagrangian strain $E_{ij} = (f_{ij} - \epsilon_{ij})/2$, 
where the radar metric reduces to 
\be
	f_{ij} = \Xup{a}{i} g_{ab}  \Xup{b}{j} 
\ee
in the $c\to\infty$ limit. 
Equations (\ref{NReoms}) are the equations of motion for a nonrelativistic elastic body. 

The nonrelativistic equations  can also be obtained 
from the $c\to\infty$ limit of the action (\ref{matteractioningauge}). 
Inserting the appropriate factors of $c$ and using the energy density (\ref{NRenergydensity}) and lapse (\ref{NRlapse}), we find 
\be
	S[X] = \int_{t_i}^{t_f}dt \int_{\cal S} d^3\zeta \, \sqrt{\epsilon} \left[ \frac{1}{2} \rho_0 \dot X^a \dot X_a  - W(E) - \rho_0 \Phi \right]
\ee
in the limit $c\to\infty$. Note that the additive constant $-\int dt \int d^3\zeta \sqrt{\epsilon} \, \rho_0 c^2 $ has been dropped from the action. 
Extremization of this action  gives the result (\ref{NReoms}), as well as the natural boundary condition 
\be
	P^i_a n_i \bigr|_{\partial {\cal S}} = 0 
\ee
on the boundary of matter space. 

For the nonrelativistic elastic body, we can define  the {\em Cauchy stress tensor} (true stress tensor) $\sigma^{ab}$  with dimensions $[\sigma^{ab}] = M/(LT^2)$. 
That is, $\sigma^{ab} n_b$ is the force per unit  of deformed area acting across a surface with unit normal $n_a$ in the deformed body. 
The Cauchy stress is related to the second Piola--Kirchhoff stress $S^{ij}$ by 
\be
	\sigma^{ab} = \frac{1}{J} \Xup{a}{i} S^{ij}\Xup{b}{j} \ ,
\ee
with $J = \sqrt{f}/\sqrt{\epsilon}$. 
In terms of the first Piola--Kirchhoff stress tensor, $\sigma^{ab} =  \Xup{(a}{i} g^{b)c} P^i_c/J$. 

\section{Isotropic Hyperelastic Models}\label{modelsection}
The energy density $\rho(E)$ is a scalar on matter space. 
For isotropic hyperelastic materials, $\rho$ depends only on the Lagrangian strain $E_{ij}$ and relaxed metric $\epsilon_{ij}$. 
For example, the Saint Venant--Kirchhoff model is defined by the potential energy density
\be
	W(E) = \frac{\lambda}{2} (\epsilon^{ij} E_{ij})^2 + \mu (\epsilon^{ik}\epsilon^{j\ell} E_{ij} E_{k\ell} ) \ ,
\ee
where $\epsilon^{ij}$ is the inverse of the relaxed metric $\epsilon_{ij}$ on ${\cal S}$. Here,  $\lambda$ and $\mu$ are the 
{\em Lam\'e constants} with dimensions $[\lambda] = [\mu] = M/(LT^2)$. 
 The second Piola--Kirchhoff stress tensor $S^{ij} = \partial W/\partial E_{ij}$ 
for the Saint Venant--Kirchhoff model is
\be
	S^{ij} = \lambda (\epsilon^{k\ell} E_{k\ell} ) \epsilon^{ij} + 2\mu \epsilon^{ik} \epsilon^{j\ell}E_{k\ell}  \ .
\ee
Note that $S^{ij}$ is a linear function of $E_{ij}$; this is not physically realistic for large stress. 

Isotropic models of hyperelastic materials are often defined in terms of the type $1\choose 1$ matter space tensor
$\epsilon^{ik} f_{kj} $
Recall that $f_{ij}$ is the right Cauchy--Green deformation tensor, which we refer to as the radar metric. The 
scalars built from  $\epsilon^{ik} f_{kj} $ are 
\bse\bea
	I_1 & \equiv & \epsilon^{ij} f_{ij} \ ,\\
	I_2 & \equiv & \frac{1}{2} \left[  (\epsilon^{ij}f_{ij} )^2 - \epsilon^{ik} \epsilon^{j\ell} f_{ij} f_{k\ell} \right] \ , \\
	I_3 & \equiv & \det(\epsilon^{ik} f_{kj}) = f/\epsilon \ ,
\eea\ese
where $f = \det(f_{ij})$ and $\epsilon = \det(\epsilon_{ij})$.   
These are the first, second and third {\em stress invariants}. Note that 
\be
	I_3 = J^2
\ee
with $J = \sqrt{f}/\sqrt{\epsilon}$, as defined in Eq.~(\ref{Jdefinition}). 

With the notation above, the Saint Venant--Kirchhoff model becomes 
\be
	W(E) = \frac{1}{8} (\lambda + 2\mu) (I_1 - 3)^2 + \mu(I_1 - 3) - \frac{\mu}{2} (I_2 - 3) \ .
\ee
Another common model for an isotropic, hyperelastic body is the Mooney--Rivlin material (\cite{Mooney}), defined by
\be
	W(E) = \frac{\mu_1}{2} (\bar I_1 - 3) + \frac{\mu_2}{2}(\bar I_2 - 3) + \frac{\kappa}{2} (J-1)^2 \ ,
\ee
where $\bar I_1 = I_1/J^{2/3}$ and $\bar I_2 = I_2/J^{4/3}$. 
For small deformations, the material parameter $\kappa$ coincides with the bulk modulus and $\mu_1 + \mu_2$ 
coincides with the shear modulus. A special case of the Mooney--Rivlin model 
is the neo--Hookean model, in which $\mu_2 = 0$. 

It is common practice to define a hyperelastic model in terms of the {\em principal stretches}, denoted $\lambda_1$, $\lambda_2$ 
and $\lambda_3$. These are defined as the square roots of the eigenvalues of $\epsilon^{ik} f_{kj}$.\footnote{The eigenvalues of $\epsilon^{ik} f_{kj}$
are positive:  $f^{ij} = \Zdown{i}{\mu} g^{\mu\nu} \Zdown{j}{\nu}$ is a positive definite matrix since  $\Zdown{i}{\mu}$ are spacelike vectors; $f_{ij}$ is positive definite since 
it is the inverse of the symmetric positive definite matrix $f^{ij}$; the eigenvalue equation $\epsilon^{ik}f_{kj} v^j = \lambda v^i$ implies 
$v^k f_{kj} v^j = \lambda v^k \epsilon_{kj} v^j$;  since $f_{ij}$ and $\epsilon_{ij}$ are both positive definite it follows that the eigenvalues $\lambda$ are positive.}  
In terms of the principal stretches, the stress invariants can be written as 
\bse\bea
	I_1 & = & \lambda_1^2 + \lambda_2^2 + \lambda_3^2 \ ,\\
	I_2 & = & \lambda_1^2\lambda_2^2 + \lambda_2^2 \lambda_3^2 + \lambda_3^2 \lambda_1^2 \ ,\\
	I_3 & = & \lambda_1^2 \lambda_2^2 \lambda_3^2 \ .
\eea\ese
As an example, the potential energy for the Ogden model \cite{Ogden} is
\be
	W(E)  = \sum_{p=1}^N \frac{\mu_p}{\alpha_p} \biggl( \lambda_1^{\alpha_p} + \lambda_2^{\alpha_p} + \lambda_3^{\alpha_p} - 3 \biggr) 
\ee
where $\mu_p$ and $\alpha_p$  are material parameters. This model is used to describe rubbers, polymers and biological tissues with large stress. 

Obviously, these  constitutive models were developed for the purpose of describing ordinary materials (rubber, steel, {\em etc}) in a nonrelativistic 
setting. We can use these same models in the relativistic regime by choosing $\rho(E) = \rho_0 c^2 + W(E)$.  

Finally, we note that a perfect fluid is a special case of an elastic material in which the energy density is a function of $J$ only:  $\rho = \rho(J)$.  
In this case the second Piola--Kirchhoff stress  is 
\be
	S^{ij} \equiv \frac{\partial \rho}{\partial E_{ij}} =  2\rho'  \frac{\partial J}{\partial f_{ij}}  = J\rho'  f^{ij} 
\ee
where $\rho' = \partial\rho/\partial J$, and the stress--energy--momentum tensor (\ref{semtensorfinal}) becomes 
\be
	T^{\mu\nu}(X(\lambda,\zeta)) = \frac{1}{J} \left[ \rho U^\nu U^\nu - J \rho' f^{ij} F^\mu_i F^\nu_j \right] \ .
\ee
Using the result (\ref{radarmetriconS}b) and the identity (\ref{invorthorelation}), along with the definition $F^\mu_i = f^\mu_\nu \Xup{\nu}{i}$ 
for the deformation gradient,  this simplifies to
\be
	T^{\mu\nu}(X(\lambda,\zeta)) = \frac{\rho}{J} U^\mu U^\nu - \rho' f^{\mu\nu} \ .
\ee
This is the SEM tensor for a perfect fluid with energy density $\rho/J$ and pressure $-\rho'$. Recall that $\rho$ is the energy per unit undeformed volume and $J$ is the ratio of deformed to 
undeformed volume. Thus, $\rho/J$ is the usual rest energy density for a perfect fluid.  The identification 
$P = -\rho'$  
for pressure comes from the first law of thermodynamics. Let ${\cal V} = d^3\zeta$ denote a coordinate volume in matter space ${\cal S}$ occupied 
by an element of fluid. 
Since $\rho' = \partial\rho/\partial J$, we can rewrite  $P = -\rho'$   as 
\be\label{firstlaw}
	d(\sqrt{\epsilon} \rho {\cal V}) = -P d(\sqrt{f} {\cal V}) \ .
\ee
On the left--hand side, $\sqrt{\epsilon} \rho {\cal V}$ is the energy of the fluid element. On the right--hand 
side, $\sqrt{f} {\cal V}$ is the physical volume  occupied by the fluid element. Equation (\ref{firstlaw}) is a statement of the first law of thermodynamics 
applied to the fluid element, relating 
the change in energy to the change in volume and the pressure $P$.

\section{Acknowledgments}\label{acknowledgements}
I would like to thank  N.~Jadoo, S.P.~Loomis and I.R.~Waldstein for helpful conversations. 
\bibliography{references}
\bibliographystyle{unsrt}
\end{document}